\documentclass[12pt]{article}
\usepackage{color}
\usepackage{graphicx}
\usepackage{slashed}
\usepackage{geometry}
\begin{document}
\title{An approach to calculation of mass spectra and two-photon decays of $c\overline{c}$ mesons in the framework of Bethe-Salpeter Equation}
\author{Shashank Bhatnagar$^{1}$ and Lmenew Alemu$^{2}$}
\maketitle

$^{1}$Department of Physics, Chandigarh University, Mohali-140413,
India
\\
$^{2}$ Department of Physics, Addis Ababa University, P.O.Box 1176, Addis Ababa, Ethiopia\\

\textbf{Abstract}\\
In this work we calculate the mass spectrum of charmonium for
$1P,...,3P$ states of $0^{++}$ , $1^{++}$, as well as for
$1S,...,4S$ states of $0^{-+}$, and $1S, ...,5D$ states of
$1^{--}$ along with the two-photon decay widths of ground and
first excited state of $0^{++}$ quarkonia for the process,
$O^{++}\rightarrow \gamma\gamma$ in the framework of a QCD
motivated Bethe-Salpeter Equation. In this $4\times 4$ BSE
framework, the coupled Salpeter equations are first shown to
decouple for the confining part of interaction, under heavy-quark
approximation, and analyically solved, and later the
one-gluon-exchange interaction is perturbatively incorporated
leading to mass spectral equations for various quarkonia. The
analytic forms of wave functions obtained are used for calculation
of two-photon decay widths of $\chi_{c0}$. Our results are in
reasonable agreement with data (where ever available) and other
models.



\section{Introduction}
An important role in applications of Quantum Chromodynamics (QCD)
to hadronic physics is played by charmonium ($c\overline{c}$) and
bottomonium ($b\overline{b}$), which are built up of a heavy quark
and heavy anti-quark. By definition, heavy quark has a mass $m$,
which is large in comparison to the typical hadronic scale,
$\Lambda_{QCD}$. Quarkonia are characterized by at least three
widely separated scales \cite{brambilla05}: the hard scale (the
mass $m$ of heavy quarks), the soft scale (relative momentum
$q\sim mv)$, and the ultra soft scale (typical kinetic energy,
$E\sim mv^{2}$ of heavy quark and anti-quark). The appearance of
all these scales in the dynamics of heavy quarkonium makes its
quantitative study extremely difficult. Quarkonium systems are
crucially important to improve our understanding of QCD. They
probe all the energy regions of QCD from hard region, where
perturbative QCD dominates, to the low energy region, where
non-perturbative effects dominate. Quarkonium states are thus a
unique laboratory where our understanding of non-perturbative QCD
may be tested.

There has a been a renewed interest in recent years in
spectroscopy of these heavy hadrons in charm and beauty sectors,
which was primarily due to experimental facilities the world over
such as BABAR, Belle, CLEO, DELPHI, BES etc.
\cite{Ecklund,babar09,belle10,cleo01,olive14}, which have been
providing accurate data on $c\overline{c}$, and $b\overline{b}$
hadrons with respect to their masses and decays. In the process
many new states have been discovered such as $\chi_{b0}(3P)$,
$\chi_{c0}(2P), X(3915), X(4260), X(4360), X(4430), X(4660)$
\cite{olive14}. The data strongly suggests that among new
resonances may be exotic four-quark states, or hybrid states with
gluonic degrees of freedom in addition to $c\overline{c}$ pair, or
loosely bound states of heavy hadrons, i.e. charmonium molecules.
Further, there are also open questions about the quantum number
assignments of some of these states such as $X(3915)$ (as to
whether it is $\chi_{c0}(2P)$ or $\chi_{c2}(2P)$
\cite{guo,olsen}). Thus charmonium offers us intriguing puzzles.

However, since the mass spectrum and the decays of all these bound
states of heavy quarks can be tested experimentally, theoretical
studies on them may throw valuable insight about the heavy quark
dynamics and lead to a deeper understanding of QCD further.
Studies on mass spectrum of these hadrons is particularly
important, since it throws light on the $Q\overline{Q}$ potential,
since the long range confinement potential can not be derived from
QCD alone. Further, though these states appear to be simple,
however, their production mechanism is still not properly
understood. These mesons are involved in a number of reactions
which are of great importance for study of
Cabibbo-Kobayashi-Maskawa (CKM) matrix and CP violation. In this
paper we also study the two-photon decays of scalar quarkonia,
$\chi_{c0}$. These decays are sensitive probes of quarkonium wave
functions.

The non-perturbative approaches, such as Effective field theory
\cite{brambilla11}, Lattice QCD \cite{mcnielle12,bali97,burch09},
Chiral perturbation theory \cite{gasser84}, QCD sum rules
\cite{shifman79,veli12}, N.R.QCD \cite{bodwin}, Bethe-Salpeter
equation (BSE)
\cite{smith69,mitra01,alkofer01,wang10,bhatnagar92,bhatnagar91,bhatnagar06,bhatnagar11,bhatnagar14,hluf17},
and potential models \cite{godfrey85,ebert13,vinodkumar16} deal
have been employed to study heavy quarkonia. Recent progress in
understanding of non-relativistic field theories make it possible
to go beyond phenomenological models, and for the first time face
the possibility of providing unified description of all aspects of
heavy quarkonium physics. This allows us to use quarkonium as a
bench mark for understanding of QCD, and for precise determination
of Standard Model parameters (e.g. heavy quark masses, QCD
coupling constant, $\alpha_s$), and for new physics searches.

All this opens up new challenges in theoretical understanding of
heavy hadrons and also provide an important tool for exploring the
structure of these simplest bound states in QCD and for studying
the non-perturbative (long distance) behavior of strong
interactions.

In present work, we do the full mass spectral problem of
charmonium  for $1P,...,3P$ states of $0^{++}$ , $1^{++}$ , as
well as for $1S,...,4S$ states of $0^{-+}$, and $1^{--}$
 along with the two-photon decay widths
of ground and first excited state of scalar quarkonia, $\chi_{c0}$
for the process, $O^{++}\rightarrow \gamma\gamma$ in the framework
of a QCD motivated Bethe-Salpeter Equation under Covariant
Instantaneous Ansatz (CIA), employing the full BS kernel
comprising both, the long-range confinement, and the short range
one-gluon-exchange (coulomb) interactions.

We do understand that in $Q\overline{Q}$ quarkonia, the
constituents are close enough to each other to warrant a more
accurate treatment of the coulomb term. Though for $b\overline{b}$
systems, the  coulomb term will be extremely dominant in
comparison to confining term, and should not be treated
perturbatively. However, seeing our mass spectral results for
$c\overline{c}$ systems, it may not be so unreasonable to treat
the coulomb term perturbatively for $c\overline{c}$ systems. This
is specially so for orbital excitations of these states, where the
centrifugal effects \cite{mitra81} ensure that the
$c-\overline{c}$ separation is large enough to feel the effect of
confining term more strongly than the coulomb term. We further
wish to state that some of earlier works
\cite{mitra81,vinodkumar08,kalinovsky05} have treated the
OGE(coulomb) term perturbatively for charmed mesons and baryons,
while some works \cite{vijayakumar13,godfrey90} did not take into
account the importance of coulomb term for heavy quarkonium
systems.

Thus, in the present paper, the coupled Salpeter equations for
scalar ($0^{++}$), and axial vector ($1^{++}$) quarkonia are first
shown to decouple for the confining part of interaction, under
heavy-quark approximation, and the analytic forms of mass spectral
equations are worked out, which are then solved in approximate
harmonic oscillator basis to obtain the unperturbed wave functions
for various states of these quarkonia. We then incorporate the
one-gluon-exchange perturbatively into the unperturbed spectral
equation, and obtain the full spectrum. The wave functions of
scalar ($0^{++}$) quarkonia, are then used to calculate their
two-photon decay widths. We further extend the mass spectral
calculations of pseudoscalar ($0^{-+}$), and vector ($1^{--}$)
quarkonia in \cite{hluf16} with the perturbative inclusion of
one-gluon-exchange effects. The approximations used in this
analytic treatment of the confining interaction are shown to be
fully under control. This work is an improvement on our earlier
work \cite{hluf16} on mass spectral problem for pseudoscalar and
vector states of quarkonia on lines of some of the earlier works
\cite{koll,babutsidze,olsson}, where we used only the confining
interaction.

A quarkonium state is classified by quantum numbers, $J^{PC}$,
where $J=L+S$, parity, $P=(-1)^{L+1}$, and charge conjugation,
$C=(-1)^{L+S}$. With this classification, while the lowest states,
$l=0$ are present in $0^{-+}$ (pseudoscalar), the lowest state
($l=0$), and the second orbitally excited $(l=2)$ state are
present in $1^{--}$ (vector) quarkonia. However, the first
orbitally excited ($l=1$) states are present in $0^{++}$ (scalar)
and $1^{++}$ (axial-vector) quarkonia. The same holds true for
their radial excitations.

This paper is organized as follows: Section 2, deals with the mass
spectral calculations of ground and excited states of $0^{++}$
quarkonia. Section 3 deals with the mass spectral calculation of
ground and excited states of $0^{-+}$, and $1^{--}$ quarkonia,
while Section 4 deals with the mass spectral calculations of
ground and excited states of $1^{++}$ quarkonia. Section 5 deals
with the calculation of two-photon decay widths of $\chi_{c0}$.
Section 6 deals with numerical results and discussions.

\section{Mass spectra of scalar quarkonia}
In the center of mass frame, where $q_\mu=(\hat{q}, i0)$, we can
write the general decomposition of the instantaneous BS wave
function for scalar mesons $(J^{pc}=0^{++})$, of dimensionality
$M$ as
\begin{equation}
  \psi(\hat{q})=Mf_{1}(\hat{q})-i\slashed{P}f_{2}(\hat{q})-i\slashed{\hat{q}}f_{3}(\hat{q})-\frac{2\slashed{P}\slashed{\hat{q}}}{M}f_{4}(\hat{q}).
  \label{sub 3053}
\end{equation}
where it can be shown by use of a power counting rule proposed in
\cite{bhatnagar06,bhatnagar14} that the Dirac structures
associated with the amplitudes $f_{1}$ and $f_{2}$ are leading,
and will contribute maximum to calculation of any scalar meson
observable, while those associated with $f_{3}$, and $f_{4}$ are
sub-leading. We now use the two constraint equations
$\psi^{+-}(\hat{q})=\psi^{-+}=0$ in the 3D Salpeter equations,
Eq.(16) of \cite{hluf16}, to reduce the four scalar functions into
two independent scalar wave functions. For equal mass system,
these two equations are reduced to
\begin{eqnarray}
&&\nonumber f_1(\hat{q})=\frac{-\hat{q}^{2}f_3(\hat{q})}{Mm};\\&&
 f_2(\hat{q})=0.
\end{eqnarray}
Applying the above constraint conditions in Eq.(2) to wave
function in Eq.(1), we rewrite the relativistic wave function of
the state $(0^{++})$ in the form
\begin{equation}
 \psi(\hat{q})=\bigg[\frac{-\hat{q}^{2}}{m}-i\slashed{\hat{q}}\bigg]f_3(\hat{q})-\frac{2\slashed{P}\slashed{\hat{q}}}{M}f_4(\hat{q}).
 \label{sub 3069}
\end{equation}
Here, it is to be noted that by use of the above constraint
equations, we have reexpressed $f_{1}$ in terms of $f_{3}$. Thus,
the Instantaneous BS wave function of $(0^{++})$ state is
determined by only two independent functions, $f_{3}$ and $f_{4}$.
Putting the wave function in Eq.(3) above, along with the
projection operators defined in Eq.(13) of \cite{hluf16}, into the
first two Salpeter equations, Eq.(16) of \cite{hluf16}, and by
taking the trace on both sides, we obtain the equations:

\begin{eqnarray}
&&\nonumber (M-2\omega)\bigg[f_3(\hat{q})+\frac{2mf_4(\hat{q})}{\omega}\bigg]=\frac{1}{\omega^2\hat{q}^2}\int\frac{d^3\hat{q}}{(2\pi)^3}K(\hat{q},
\hat{q}')\bigg[\hat{q}^2\hat{q}'^2f_3(\hat{q}')-m^2\hat{q}.\hat{q}'f_3(\hat{q}')-2m\omega\hat{q}.\hat{q}'f_4(\hat{q}')\bigg]\\&&
(M+2\omega)\bigg[f_3(\hat{q})-\frac{2mf_4(\hat{q})}{\omega}\bigg]=\frac{1}{\omega^2\hat{q}^2}\int\frac{d^3\hat{q}}{(2\pi)^3}K(\hat{q}, \hat{q}')\bigg[-\hat{q}^2\hat{q}'^2f_3(\hat{q}')+m^2\hat{q}.\hat{q}'f_3(\hat{q}')-2m\omega\hat{q}.\hat{q}'f_4(\hat{q}')\bigg]
\label{sub 3073}
 \end{eqnarray}

Solution of these equations needs information about the BS kernel,
$K(\hat{q}, \hat{q}')$ \cite{bhatnagar14,hluf16}, which is taken
to be one-gluon exchange like as regards the colour
($\frac{1}{4}\vec{\lambda}_1.\vec{\lambda}_2$) and spin
($\gamma_\mu\ \bigotimes \gamma_\mu$) dependence, and has a scalar
part $V(\hat{q}, \hat{q}')$, written as,

\begin{equation}
V(\hat{q},\hat{q}')=\frac{4\pi\alpha_{s}}{(q-q')^{2}}+
\frac{3}{4}\omega^{2}_{q\bar{q}}\int d^{3}\vec{r}[\kappa
r^{2}-\frac{C_{0}}{\omega_{0}^{2}}]
e^{i(\hat{q}-\hat{q}').\vec{r}}=V_{OGE}+V_{c}.
\end{equation}

Thus, the scalar part $V(\hat{q},\hat{q}')$ of the kernel involves
both the OGE term $V_{OGE}$, arising from the one-gluon exchange,
as well as the confining term, $V_{c}$. We first ignore the
$V_{OGE}$ term, and work only with the confining part, $V_{c}$ in
Eq.(5), and write Eqs.(4) as,

\begin{eqnarray}
 &&\nonumber (M-2\omega)\bigg[f_3(\hat{q})+\frac{2mf_4(\hat{q})}{\omega}\bigg]=\frac{\Theta_s}{\omega^2\hat{q}^2}\int \frac{d^3\hat{q}'}{(2\pi)^3}V_{c}(\hat{q}, \hat{q}')\bigg[\hat{q}^2\hat{q}'^2f_3(\hat{q}')-m^2\hat{q}.\hat{q}'f_3(\hat{q}')-2m\omega\hat{q}.\hat{q}'f_4(\hat{q}')\bigg]\\&&
 (M+2\omega)\bigg[f_3(\hat{q})-\frac{2mf_4(\hat{q})}{\omega}\bigg]=\frac{\Theta_s}{\omega^2\hat{q}^2}\int \frac{d^3\hat{q}'}{(2\pi)^3}V_{c}(\hat{q}, \hat{q}')\bigg[-\hat{q}^2\hat{q}'^2f_3(\hat{q}')+m^2\hat{q}.\hat{q}'f_3(\hat{q}')-2m\omega\hat{q}.\hat{q}'f_4(\hat{q}')\bigg],
 \label{sub 3077}
 \end{eqnarray}

where the spin dependence of the interaction is contained in the
factor,  $\Theta_{S}=\gamma_{\mu}\psi(\widehat{q})\gamma_{\mu}$.
The scalar part of the confining potential (that involves the
colour factor
$\frac{1}{2}\vec{\lambda}_{1}.\frac{1}{2}\vec{\lambda}_{2}=
-\frac{4}{3}$), is taken to be \cite{hluf16}
$V_{c}(\hat{q},\hat{q'})=\overline{V}_{c}
\delta^{3}(\hat{q}-\hat{q}')$, where
$\overline{V}_{c}(\hat{q},\hat{q'})=\omega_{q\bar{q}}^{2}(2\pi)^{3}\big[\kappa\vec{\nabla}_{\hat{q}}^{2}+\frac{C_{0}}{\omega_{0}^{2}}\big]$,
and
$\kappa=(1-A_{0}M^{2}\overrightarrow{\nabla}^{2}_{\hat{q}})^{-1/2}$.
Here $\overline{V_{c}}$ is the part of $V_{c}$ without the delta
function.  To handle these equations, we first integrate these
equations over $d^{3}\widehat{q}'$, performing the delta function
integration that arises due to presence of
$V_{c}(\hat{q},\hat{q'})$ in the integrand, and get two coupled
algebraic equations with $\overline{V}(\widehat{q})$ on RHS. To
decouple them, we first add them. Then we subtract the second
equation from the first equation, and get two algebraic equations
which are still coupled. Then from one of the two equations so
obtained, we eliminate $f_{3}(\hat{q})$ in terms of
$f_{4}(\hat{q})$, and plug this expression for $f_{3}(\hat{q})$ in
the second equation of the coupled set so obtained, to get a
decoupled equation in $f_{4}(\hat{q})$. Similarly, we eliminate
$f_{4}(\hat{q})$ from the second equation of the set of coupled
algebraic equations in terms of $f_{3}(\hat{q})$, and plug it into
the first equation to get a decoupled equation entirely in
$f_{3}(\hat{q})$. Thus, we get two identical decoupled equations,
one entirely in $f_{3}(\hat{q})$, and the other that is entirely
in, $f_{4}(\hat{q})$. The calculation up to this point is without
any approximation. However, we notice that if we employ the
approximation, $\omega\approx m$ on RHS of the two algebraic
equations {\footnote{In principle, we should solve the decoupled
algebraic equations numerically. However, this would not give
explicit dependence of the mass spectra on the principal quantum
number N, and nor will this give the explicit algebraic forms of
wave functions that can be employed to do analytic calculations of
various transition amplitudes for different processes. Our
approach may lead to a little loss of numerical accuracy, but it
does lead to a much deeper understanding of the mass spectral
problem. The approximations used by us have been shown to be
totally under control. The plots of our algebraic forms of wave
functions for scalar, pseudoscalar, vector and axial vector
quarkonia are very much similar to the corresponding plots of wave
functions in \cite{wang10} obtained by purely numerical methods,
which validates our approach.}}(this is justified since in the
confinement region, the relative momentum between heavy quarks in
the bound state can be considered small, since the heavy-quark is
expected to move with non-relativistic speeds (see
Ref.\cite{shi14}), and these quarks can be treated as almost on
mass shell), these decoupled equations can be expressed as:

\begin{eqnarray}
&&\nonumber
\bigg[\frac{M^{2}}{4}-m^{2}-\widehat{q}^{2}\bigg]f_{3}(\hat{q})=-m\Theta_{s}\overline{V}_{c}f_{3}(\hat{q})+\frac{\Theta_{s}^{2}\overline{V}_{c}^{2}f_{3}(\widehat{q})}{4}\\&&
\bigg[\frac{M^{2}}{4}-m^{2}-\widehat{q}^{2}\bigg]f_{4}(\hat{q})=-m\Theta_{s}\overline{V}_{c}f_{4}(\hat{q})+\frac{\Theta_{s}^{2}\overline{V}_{c}^{2}f_{4}(\widehat{q})}{4},
\label{sub 3109}
\end{eqnarray}
We see that we get two identical decoupled equations which
resemble the harmonic oscillator equations, but for the term
involving $\overline{V}_{c}^{2}$ on the right side of these
equations.

We wish to mention that in recent studies on mass spectra of
pseudoscalar and vector quarkonia \cite{hluf16}, it was seen that
good agreement with data on masses and various decay
constants/decay widths of ground and excited states of $\eta_{c},
\eta_{b}, J/\Psi$, and $\Upsilon$ is obtained for input
parameters, $C_{0}=0.175$, $\omega_{0}=0.160$ GeV.,
$\Lambda=0.200$ GeV., and $A_{0}=0.01$, along with the input quark
masses $m_c=1.490$ GeV. With these numerical values of input
parameters, we try to determine the numerical values of
$\Omega_{S}=m\Theta_{S}\omega^{2}_{q\bar{q}}$, and $\Omega_{S}'=
\frac{\Theta_{S}^2}{4}\omega^{4}_{q\bar{q}}$ associated with the
terms involving $\overline{V}_{c}$ and $\overline{V}_{c}^{2}$
respectively, for scalar mesons $\chi_{c}$ and $\chi_{b}$ in RHS
of Eqs.(12), and their percentage ratio in Table 1 below, where it
can be seen that $\omega_{q\bar{q}}^4\ll\omega_{q\bar{q}}^2$ and
hence the second term on RHS of Eqs.(7) contributes $< 1\%$, than
the first term on RHS of this equation for  $c\overline{c}$, so
that it can be dropped.
\bigskip

\begin{table}[h]
  \begin{center}
\begin{tabular}{llll}
  \hline
     &$\Omega_{S}$&$\Omega_{S}'$&$\frac{\Omega_{S}'}{\Omega_{S}}\%$\\\hline
    $\chi_{c0}$&0.0558&0.000356&0.638\\\
    \end{tabular}
\end{center}
\caption{Numerical values of coefficients,
$\Omega_{S}=m\Theta_{S}\omega^{2}_{q\bar{q}}$, and $\Omega_{S}'=
\frac{\Theta_{S}^2}{4}\omega^{4}_{q\bar{q}}$ associated with the
terms involving $\overline{V}_{c}$ and $\overline{V}_{c}^{2}$
respectively, for scalar mesons $\chi_{c}$ in RHS of Eqs.(7), and
their percentage ratio for the input parameters of our model
mentioned above.}
\end{table}
\bigskip

Thus the RHS of both equations in Eq.(7) has only the term,
$-m\Theta_{s}\overline{V}_{c}f_{3,4}(\widehat{q})$. Now, putting
the spatial part, $\overline{V}_{c}$ in Eq.(7), the wave functions
$f_{3}$, and $f_{4}$ then satisfy identical 3D BSE for equal mass
heavy scalar mesons:

\begin{eqnarray}
  &&\nonumber \bigg[\frac{M^{2}}{4}-m^{2}-\hat{q}^2\bigg]f_{3}(\hat{q})=-m\Theta_{s}\omega_{q\bar{q}}^{2}\big[{\kappa\vec{\nabla}_{\hat{q}}}^{2}+\frac{C_{0}}{\omega_{0}^{2}}\big]f_{3}(\hat{q})\\&&
  \bigg[\frac{M^{2}}{4}-m^{2}-\hat{q}^2\bigg]f_{4}(\hat{q})=-m\Theta_{s}\omega_{q\bar{q}}^{2}\big[{\kappa\vec{\nabla}_{\hat{q}}}^{2}+\frac{C_{0}}{\omega_{0}^{2}}\big]f_{4}(\hat{q})
 \label{sub 3110}
 \end{eqnarray}

where we have used
$\overline{V}(\hat{q})=(2\pi)^{3}\omega^{2}_{q\bar{q}}[\kappa\overrightarrow{\nabla}^{2}_{\hat{q}}
+\frac{C_{0}} {\omega^{2}_{0}}]$ \cite{hluf16}, with
$\kappa=(1+2A_{0}(N+\frac{3}{2}))^{-1/2}$. Thus the solutions of
these equations, $f_{3}(\hat{q})\approx f_{4}(\hat{q}) ( =
\phi^{s}(\hat{q}))$. With the use of the above equality of
amplitudes, and reexpressing $f_{3}$ in terms of $f_{1}$, the
complete wave function $\Psi^{s}(\widehat{q})$ can be expressed
as:
\begin{equation}
\Psi^{s}(\widehat{q})=[M+i\frac{mM\slashed{\widehat{q}}}{\widehat{q}^{2}}-\frac{2\slashed{P}\slashed{q}}{M}]\phi^{s}(
\widehat{q}).
\end{equation}
 We can reduce this equation into the equation of a simple quantum mechanical $3D$-harmonic oscillator with coefficients
depending on the hadron mass $M$, and total quantum number $N$.
The wave function satisfies the 3D BSE:
 \begin{equation}
  \bigg[\frac{M^{2}}{4}-m^{2}-\hat{q}^2\bigg]\phi(\hat{q})=-m\Theta_{s}\omega_{q\bar{q}}^{2}\big[{\kappa\vec{\nabla}_{\hat{q}}}^{2}+\frac{C_{o}}{\omega_{o}^{2}}\big]\phi(\hat{q}).
  \label{sub 3111}
 \end{equation}

Now, with the use of leading Dirac structures, we can to a good
approximation (as in case of pseudoscalar and vector mesons
\cite{hluf16}) express $\Theta_{S}=4$. This is due to the fact
that $\gamma_\mu\psi(\widehat{q})\gamma_\mu \approx\gamma_\mu
(Mf_{1})\gamma_{\mu}=4\psi(\hat{q})$, due to $MI$ ($I$ being the
unit $4\times 4$ matrix) being the most leading Dirac structure
in the scalar meson wave function, and the terms with
$\widehat{q}^{2}/m^{2}$ have negligible contributions, in heavy
quark limit, and can be dropped. Thus the above equation can be
put in the form
\begin{equation}
  (\frac{M^2}{4}-m^{2}-\hat{q}^{2})\phi(\hat{q})=-\beta_{s}^{4}\big[\kappa\vec{\nabla}_{\hat{q}}^{2}+\frac{C_{0}}{\omega_{0}^{2}}\big]\phi(\hat{q}).
\label{sub 3112}
  \end{equation}

which can in turn be expressed as,
\begin{equation}
E_{s}\phi_{s}=[-\beta_{s}^{4}
\overrightarrow{\nabla}_{\widehat{q}}^{2}+\widehat{q}^{2}]\phi_{s}(\widehat{q}),
 \label{sub 3114}
\end{equation}

where,
$\beta_{s}=(\frac{4m\omega_{q\bar{q}}^{2}}{\sqrt{1+2A_{0}(N+3/2)}})^{1/4}$,
and with total energy of the system expressed as,
\begin{equation}
  E=\frac{M^2}{4}-m^{2}+\frac{\beta_{s}^{4}C_{0}}{\omega_{0}^{2}}\sqrt{1+2A_{0}(N+3/2)}.
  \label{sub 3115}
\end{equation}
Putting the expression for the laplacian operator in spherical
coordinates, we get
\begin{equation}
  \frac{d\phi(\hat{q})}{d\hat{q}^{2}}+\frac{2}{\hat{q}}\frac{d\phi(\hat{q})}{d\hat{q}}-\frac{l(l+1)\phi(\hat{q})}{\hat{q}^{2}}+(\frac{E}{\beta_{s}^{4}}-\frac{\hat{q}^{2}}{\beta_{s}^{4}})\phi(\hat{q})=0
\label{sub 3116}
  \end{equation}
where $l$ is the orbital quantum number with the values $l=0, 1,
2, 3 ...$, corresponding to $S, P, D,...$ wave states
respectively. This is a 3D harmonic oscillator equation, whose
solutions can be found by using power series method. Assuming the
form of the solutions of this equation as,
$\phi(\hat{q})=h(\hat{q})e^{-\frac{\hat{q}^{2}}{2\beta_{s}^{2}}}$,
the above equation can be expressed as,

\begin{equation}
  h''(\hat{q})+(\frac{2}{\hat{q}}-\frac{2\hat{q}}{\beta_{s}^{2}})h'(\hat{q})+\bigg(\frac{E}{\beta_{s}^{4}}-\frac{3}{\beta_{s}^{2}}-\frac{l(l+1)}{\hat{q}^{2}}\bigg)h(\hat{q}).
\label{sub 3119}
\end{equation}

The eigen values of this equation can be obtained using the power
series method as,

\begin{equation}
E_{N}=2\beta_{s}^{2}\big[N+\frac{3}{2}\big]; N=2n+l, \label{sub
3137}
\end{equation}
with $l=1$. Thus, to each value of $n=0, 1, 2, 3, ...$ will
correspond a polynomial $h(\hat{q})$ of order $2n+1$ in $\hat{q}$,
that are obtained as solutions of Eq.(15). The odd parity
normalized wave functions $\phi(\hat{q})$ thus derived are:

\begin{eqnarray}
&&\nonumber\phi_s(1p,
\hat{q})=(\frac{2}{3})^{1/2}\frac{1}{\pi^{3/4}\beta_s^{5/2}}\hat{q}e^{-\frac{\hat{q}^{2}}{2\beta_s^{2}}},\\&&
\nonumber \phi_s(2p,
\hat{q})=(\frac{5}{3})^{1/2}\frac{1}{\pi^{3/4}\beta_s^{5/2}}
\hat{q}\bigg[1-\frac{2}{5}\frac{\hat{q}^{2}}{\beta_s^2}
\bigg]e^{-\frac{\hat{q}^{2}}{2\beta_s^{2}}},\\&& \phi_s(3p,
\hat{q})=(\frac{70}{171})^{1/2}\frac{1}{\pi^{3/4}\beta_s^{5/2}}
\hat{q}\bigg[1-\frac{2\hat{q}^{2}}{5\beta_s^{4}}+
\frac{4\hat{q}^{4}}{35\beta_s^4}
\bigg]e^{-\frac{\hat{q}^{2}}{2\beta_s^{2}}} \label{sub 3159}
\end{eqnarray}

The plots of wave functions for scalar quarkonia, $\chi_{c0}$ are
given in Fig. 1 below.

\begin{figure}[h]
\centering
\includegraphics[width=16cm]{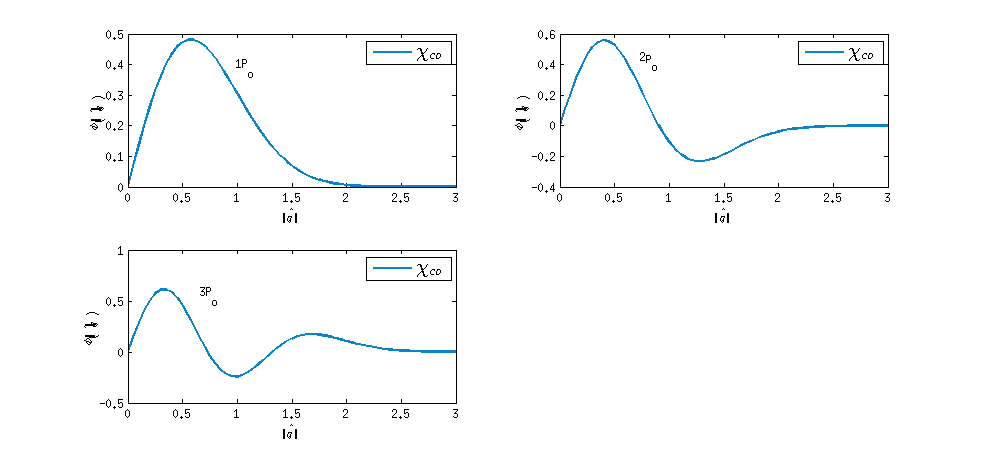}
\caption{Plots of wave functions for scalar ($0^{++}$) quarkonia
$\chi_{c0}$ Vs $\widehat{q}$ (in Gev.) for the states $1P, 2P$ and
$3P$.}
\end{figure}

The mass spectrum of ground and excited states for equal mass
heavy scalar ($0^{++}$) mesons is written as:
\begin{equation}
\frac{1}{2\beta_{s}^{2}}(\frac{M^{2}}{4}-m^{2}+\frac{C_{0}\beta_{s}^{4}}{\omega^{2}_{0}}
 \sqrt{1+2A_{0}(N+\frac{3}{2})})=N+\frac{3}{2}; N=2n+l; n=0,1,2,...,
\end{equation}

where the orbital quantum number, $l=1$. Now, treating the mass
spectral equation, Eq.(12) as the unperturbed equation, with the
unperturbed wave functions, $\phi_{S}(nP,\widehat{q})$ for scalar
mesons in Eq.(17), we now incorporate the OGE (coulomb) term in
this equation.

Then the above mass spectral equation can be written as:
\begin{equation}
E_{S}\phi_{S}(\hat{q})=[-\beta_{S}^{4}\overrightarrow{\nabla}^{2}_{\hat{q}}+\hat{q}^{2}+V_{coul}^{S}]\phi_{S}(\hat{q}),
\end{equation}

Treating the coulomb term as a perturbation to the unperturbed
mass spectral equation, we can write the complete mass spectra for
the ground and excited states for equal mass heavy scalar
($0^{++}$) mesons using first order perturbation theory as:

\begin{equation}
\frac{1}{2\beta_{S}^2}\{\frac{M^{2}}{4}-m^{2}+\frac{\beta_{S}^{4}C_{0}}{\omega_{0}^{2}}\sqrt{1+2A_{0}(N+\frac{3}{2})}\}+
\gamma <V_{coul}^{S}>=(N+\frac{3}{2}); N=2n+l; n=0,1,2....,
\end{equation}

with $l=1$, where $<V_{coul}^{S}>$  is the expectation value of
$V_{coul}^{S}$ between the unperturbed states of given quantum
numbers $n$ (with $l=1$) for scalar mesons, and has been weighted
by a factor of
$\gamma=\frac{C_{0}^2\beta_{S}^{4}}{(m_{1}+m_{2})^{2}}$ to have
the coulomb term dimensionally consistent with the harmonic term.
Its expectation values for the $1P, 2P$, and $3P$ states are:
\begin{eqnarray}
&&\nonumber
<1P|V_{coul}^{S}|1P>=-\frac{128\pi\alpha_{s}}{9\beta_{S}^{2}},\\&&
\nonumber
<2P|V_{coul}^{S}|2P>=-\frac{64\pi\alpha_{s}}{18\beta_{S}^{2}},\\&&
<3P|V_{coul}^{S}|3P>=-\frac{3712\pi\alpha_{s}}{213\beta_{S}^{2}}
\end{eqnarray}

\begin{table}[hhhhh]
  \begin{center}
\begin{tabular}{|l|l|l|l|l|l|}
  \hline
   & BSE-CIA &Expt \cite{olive14}&Pot.Model\cite{godfrey85} &BSE\cite{wang10} &RQM\cite{ebert13}\\
   \hline
  $M_{\chi c0}(1p_0)$ & 3.4186& 3.4140$\pm $0.0003 & 3.440 &  & 3.413 \\
  \hline
  $M_{\chi c0}(2p_0)$ &4.1804 &  &3.9200  &3.8368     &3.8700  \\
  \hline
 $ M_{\chi c0}(3p_0)$ &5.0323&  &  &4.1401  & 4.3010 \\
  \hline

  \end{tabular}
\caption{Mass spectrum of ground and excited states of $\chi_{c0}$
 with quantum numbers $J^{PC}=0^{++}$ in GeV. units
with the above set of parameters.}
\end{center}
\end{table}

From the mass spectral equation, one can see that, the mass
spectra depends not only on the principal quantum number $N$, but
also the orbital quantum number $l$. We are now in a position to
calculate the numerical values for mass spectral of heavy equal
mass scalar meson with the input parameters of our model. The
results of mass spectral predictions of heavy equal mass scalar
mesons for both ground and excited states with the above set of
parameters is given in table 2.

We now derive the mass spectral equations with the incorporation
of the OGE (Coulomb) term for pseudoscalar and vector quarkonia,
and obtain their solutions in the next section (the preliminary
calculations using only the confining part of interaction were
done in \cite{hluf16}).

\section{Mass spectral equation for pseudoscalar ($0^{-+}$), and
vector ($1^{--}$) quarkonia} For pseudoscalar (P), and vector (V)
quarkonia, the general decomposition of instantaneous BS wave
function of dimensionality $M$ in the center of mass frame is
given in Eqs. (18), and (25) respectively in \cite{hluf16}.
Putting this wave function in Eqs.(18)(for P-mesons), or Eq.(25)
(for V-mesons) into the Salpeter equations leads to two coupled
equations in leading amplitudes ($\phi_1$, and $\phi_2$) for
P-mesons, and ($\chi_1$, and $\chi_2$) for V-mesons. Decoupling
them in the heavy-quark limit, leads to Eqs.(37) (for both P and
V-mesons) of \cite{hluf16} given as,

\begin{equation}
E_{P,V}\phi_{s}=[-\beta_{P,V}^{4}
\overrightarrow{\nabla}_{\widehat{q}}^{2}+\widehat{q}^{2}]\phi_{P,V}(\widehat{q}),
\end{equation}
with non-perturbative energy eigen functions for $l=0 (S)$, and
for $l=2 (D)$ states obtained as solutions of above spectral
equation in approximate harmonic oscillator basis for pseudoscalar
quarkonia (for states $1S,...,4S$) and vector quarkonia (for
states $1S,...,3D$) as Eq.(41) in \cite{hluf16}, for the
perturbative calculation of the short ranged one-gluon-exchange
interaction shown later in this section.

The plots of unperturbed wave functions for pseudoscalar and
vector quarkonia are given in \cite{hluf16}. The unperturbed mass
spectra is expressed as (see\cite{hluf16}),
\begin{equation}
 [\frac{M^2}{4}-m^{2}+\frac{\beta_{P,V}^{4}C_{0}}{\omega_{0}^{2}}\sqrt{1+2A_{0}(N+3/2)}]=2\beta_{P,V}^{2}(N+\frac{3}{2});
 N=2n+l; n=0,1,2,...
 \end{equation}
The mass spectra of vector charmonium and bottomonium states using
the above spectral equation was found to have degenerate $S$, and
$D$ states \cite{hluf16}. However with the incorporation of the
OGE (Coulomb) term the mass spectral equation can be written as:
\begin{equation}
E_{P,V}\phi_{P,V}(\hat{q})=[-\beta_{P,V}^{4}\overrightarrow{\nabla}^{2}_{\hat{q}}+\hat{q}^{2}+V_{coul}^{P,V}]\phi_{P,V}(\hat{q}),
\end{equation}
Now, treating the coulomb term as a perturbation to the
unperturbed mass spectral equation, Eq.(22), and treating the wave
functions $\phi_{P,V}(\widehat{q})$ in Eq.(41) of \cite{hluf16},
as the unperturbed wave functions, we can write the complete mass
spectra of ground (1S) and excited states for equal mass heavy
pseudoscalar ($0^{-+}$) and vector ($1^{--}$) mesons respectively
using the first order degenerate perturbation theory as:

\begin{equation}
\frac{1}{2\beta_{P,V}^2}\{\frac{M^{2}}{4}-m^{2}+\frac{\beta_{P,V}^{4}C_{0}}{\omega_{0}^{2}}\sqrt{1+2A_{0}(N+\frac{3}{2})}\}+
\gamma <V_{coul}^{P,V}>=(N+\frac{3}{2}); N=2n+l; n=0,1,2....,
\end{equation}

where $<V_{coul}^{P,V}>$ (has again been weighted by a factor of
$\gamma_{P,V}=\frac{C_{0}^{2}\beta_{P,V}^{4}}{(m_{1}+m_{2})^{2}}$,
as in scalar ($0^{++}$) quarkonia in previous section) is the
matrix element of $V_{coul}^{P,V}$ between unperturbed states in
Eq.(41) of \cite{hluf16} of given quantum numbers $n$, and $l$,
(with $n=0,1,2,3,...$, and $l= 0$ for pseudoscalar mesons, while
with $l= 0,2$ for vector mesons). It is to be noted that,
$V_{coul}^{P,V}$ connects only the equal parity states with the
same values of quantum numbers $n$. The only non-vanishing matrix
elements of the perturbation between states with the given quantum
numbers $n$ and $l$ are listed below:
\begin{eqnarray}
&&\nonumber
<nS|V_{coul}^{P}|nS>=\frac{\pi\alpha_{s}}{12}\frac{1}{\beta_{P}^{2}}\\&&
\nonumber
<nS|V_{coul}^{V}|nS>=\frac{\pi\alpha_{s}}{24}\frac{1}{\beta_{V}^{2}}\\&&
<nD|V_{coul}^{V}|nD>=\frac{\pi\alpha_{s}}{24}\frac{1}{5\beta_{V}^{2}},
\end{eqnarray}

The non-zero values of $<V_{coul}>$ given above not only lead to
the lifting up of the degeneracy between the $S$ and $D$ levels
with the same principal quantum number $N$ in vector quarkonia,
but also leads to bringing the masses of different states of
vector and pseudoscalar quarkonia closer to data, as can be seen
from the mass spectral results for pseodoscalar ($0^{-+}$), and
vector ($1^{--}$) quarkonia, which are compared with the
experimental data \cite{olive14}, and other models for each state
(where ever available), as given in Tables 3 and 4 respectively
as:

\begin{table}[h]
\begin{center}
\begin{tabular}{lllllll}
  \hline
    &BSE - CIA &Expt.\cite{olive14}&Pot.
    Model\cite{bhagyesh11}&
   QCD sum rule\cite{veli12}&Lattice QCD\cite{burch09}&\cite{ebert13} \\\hline
    $M_{\eta_{c}(1S)}$& 2.9822 & 2.983$\pm$0.0007&2.980 & 3.11$\pm$0.52 &3.292&2.981 \\
    $M_{\eta_{c}(2S)}$& 3.7395 & 3.639$\pm$0.0013  &3.600& &4.240&3.635\\
    $M_{\eta_{c}(3S)}$& 4.4256 &   &4.060& &&3.989\\
    $M_{\eta_{c}(4S)}$& 5.0843&    &4.4554& &&4.401\\
    \hline
     \end{tabular}
   \end{center}
   \caption{Masses of ground and radially excited
states of $\eta_c$ (in GeV.) in present calculation (BSE-CIA)
along with experimental data, and their masses in other models.}
\end{table}

\begin{table*}[tbp]
  \begin{center}
\begin{tabular}{lllllll}
  \hline
    &BSE - CIA &Expt.\cite{olive14}&Rel. Pot. Model\cite{ebert13}&Pot. Model\cite{bhagyesh11}
    &BSE\cite{wang10}&Lattice QCD\cite{kawanai15}\\\hline
    $M_{J/\psi(1S)}$&3.0897& 3.0969$\pm$ 0.000011&3.096       & 3.0969 & &3.099 \\
    $M_{\psi(2S)}$&3.7004& 3.6861$\pm$ 0.00034  &3.685& 3.6890&3.686 &3.653\\
    $M_{\psi(1D)}$&3.7878&3.773$\pm$ 0.00033&3.783  & &3.759&\\
    $M_{\psi(3S)}$&4.278&4.03$\pm$ 0.001&4.039       & 4.1407&4.065 &4.099\\
    $M_{\psi(2D)}$&4.3444&4.191$\pm$0.005&4.150 &&4.108 &\\
    $M_{\psi(4S)}$& 4.8035&4.421$\pm$0.004 &4.427&4.5320&4.344& \\
    $M_{\psi(3D)}$& 4.8617&&4.507 && 4.371  & \\
    $M_{\psi(5S)}$& 5.1391&& 4.837   &4.8841 & 4.567 &    \\
    $M_{\psi(4D)}$& 5.1449&&4.857    & &  &    \\
       \hline
     \end{tabular}
   \end{center}
   \caption{Masses of ground, radially and orbitally excited states of
heavy vector quarkonium, $J/\psi$  in BSE-CIA along with their
masses in other models and experimental data (all units are in
GeV).}
   \end{table*}

\section{Mass spectral equation for axial vector $1^{++}$ quarkonia}
The general form for the relativistic Salpeter wave function of
$^3P_1$ state with $J^{PC}=1^{++}$ can be expressed as in
\cite{smith69,alkofer01}. In the center of mass frame, we can then
write the general decomposition of the instantaneous BS wave
function for axial vector mesons $(J^{pc}=1^{++})$ of
dimensionality $M$ as:

\begin{eqnarray}
&&\nonumber \psi(\hat{q})=\gamma_5[\gamma_\mu+\frac{P_\mu \slashed{P}}{M^2}][iM g_1(\hat{q})+\slashed{P}g_2(\hat{q})-\slashed{\hat{q}}g_3(\hat{q})+2i\frac{\slashed{P}\slashed{\hat{q}}}{M}g_4(\hat{q})]\\&&
 +\gamma_5[M\hat{q_\mu}g_3(\hat{q})+2i\hat{q_\mu}\slashed{P}g_4(\hat{q})].
\end{eqnarray}

With use of our power counting rule
\cite{bhatnagar06,bhatnagar11}, it can be checked that the Dirac
structures associated with amplitudes $g_{1}$ and $g_{2}$ are
$O(M^{1})$, and are leading, and would contribute maximum to any
axial vector meson calculation. Following a similar procedure as
in the case of scalar mesons, we can write the Salpeter wave
function in terms of only two leading Dirac amplitudes $g_1$, and
$g_2$. Plugging this wave function together with the projection
operators into the first two Salpeter equations in Eq.(16) of
\cite{hluf16}, and taking trace on both sides and following the
steps as for the scalar meson case, we get the coupled integral
equations in the amplitudes $g_{1}$, and $g_{2}$:

\begin{eqnarray}
&&\nonumber (M-2\omega)[-\frac{2m g_{1}}{\omega}+2g_{2}]=
\Theta_{A}\int \frac{d^{3} \widehat{q}'}{(2\pi)^{3}}V_{c}(
\widehat{q}, \widehat{q}')[-\frac{2m g_{1}}{\omega}+2g_{2}],\\&&
(M-2\omega)[-\frac{2m g_{1}}{\omega}+2g_{2}]=-\Theta_{A}\int
\frac{d^{3} \widehat{q}'}{(2\pi)^{3}}V_{c}( \widehat{q},
\widehat{q}')[-\frac{2m g_{1}}{\omega}+2g_{2}].
\end{eqnarray}

To decouple these equations, we follow a similar procedure as in
the scalar meson case, and get two identical decoupled equations,
one entirely in $g_{1}(\hat{q})$, and the other that is entirely
in, $g_{2}(\hat{q})$. In the limit, $\omega\approx m$ on RHS, and
due to the fact that for axial quarkonia again,
$\omega_{q\bar{q}}^4\ll\omega_{q\bar{q}}^2$ these equations can be
expressed as:

\begin{eqnarray}
 &&\nonumber[\frac{M^2}{4}-m^2-\hat{q}^2]g_1(\hat{q})=-m\Theta_A\omega_{q\bar{q}}^2[\vec{\nabla}_{\hat{q}}^2+\frac{C_0}{\omega_0^2}]g_1(\hat{q})\\&&
 [\frac{M^2}{4}-m^2-\hat{q}^2]g_2(\hat{q})=-m\Theta_A\omega_{q\bar{q}}^2[\vec{\nabla}_{\hat{q}}^2+\frac{C_0}{\omega_0^2}]g_2(\hat{q})
\end{eqnarray}

where, it can be checked that both $g_1$ and $g_2$ satisfy the
same equation and hence we can approximately write $g_1\sim
g_2\sim \phi_A$. Thus,

\begin{equation}
 [\frac{M^2}{4}-m^2-\hat{q}^2]\phi_{A}(\hat{q})=-\Theta_{A}
 m\omega_{q\bar{q}}^2[\vec{\nabla}_{\hat{q}}^2+\frac{C_0}{\omega_0^2}]\phi_{A}(\hat{q}),
\end{equation}
where $\Theta_{A}=\gamma_{\mu}\Psi(\widehat{q})\gamma_{\mu}$. Now,
with the use of leading Dirac structures, we can again to a good
approximation express $\Theta_{A}=2$. This is due to the fact that
$\gamma_\mu\psi(\widehat{q})\gamma_\mu \approx\gamma_\mu
\gamma_{5}\gamma_{\nu}(iMg_{1})\gamma_{\mu}=2\gamma_{5}\gamma_{\nu}(iMg_{1})\approx
2\psi(\widehat{q})$. This mass spectral equation has the same form
as the mass spectral equation for scalar quarkonia, Eq. (10),
except for the value of $\Theta_{A}$, which is different from
$\Theta_{S}$. Thus the above equation can be put in a similar form
as Eq.(12) (for scalar case), except that inverse range parameter
$\beta_{s}\rightarrow \beta_{A}$, where,
$\beta_{A}=(\frac{2m\omega_{q\bar{q}}^{2}}{\sqrt{1+2A_{0}(N+3/2)}})^{1/4}$,
and the $3D$ wave function, $\phi_{s}\rightarrow \phi_{A}$.  The
mass spectral equation for $1^{++}$ would then exactly resemble
Eqs.(18) for $0^{++}$ case, and thus the unperturbed wave
functions $\phi_{A}(\widehat{q})$ for $1^{++}$ would then have the
same algebraic form as $\phi_{s}(\widehat{q})$ in Eqs.(17), but
with $\beta_{s}\rightarrow \beta_{A}$:

\begin{eqnarray}
&&\nonumber\phi_{A}(1p,
\hat{q})=(\frac{2}{3})^{1/2}\frac{1}{\pi^{3/4}\beta_{A}^{5/2}}\hat{q}e^{-\frac{\hat{q}^{2}}{2\beta_{A}^{2}}},\\&&
\nonumber \phi_A(2p,
\hat{q})=(\frac{5}{3})^{1/2}\frac{1}{\pi^{3/4}\beta_A^{5/2}}
\hat{q}\bigg[1-\frac{2}{5}\frac{\hat{q}^{2}}{\beta_A^2}
\bigg]e^{-\frac{\hat{q}^{2}}{2\beta_A^{2}}},\\&& \phi_A(3p,
\hat{q})=(\frac{70}{171})^{1/2}\frac{1}{\pi^{3/4}\beta_A^{5/2}}
\hat{q}\bigg[1-\frac{2\hat{q}^{2}}{5\beta_A^{4}}+
\frac{4\hat{q}^{4}}{35\beta_A^4}
\bigg]e^{-\frac{\hat{q}^{2}}{2\beta_A^{2}}}
\end{eqnarray}

The plots of wave functions for axial vector quarkonia are given
in Fig.2 below

\begin{figure}[h]
\centering
\includegraphics[width=16cm]{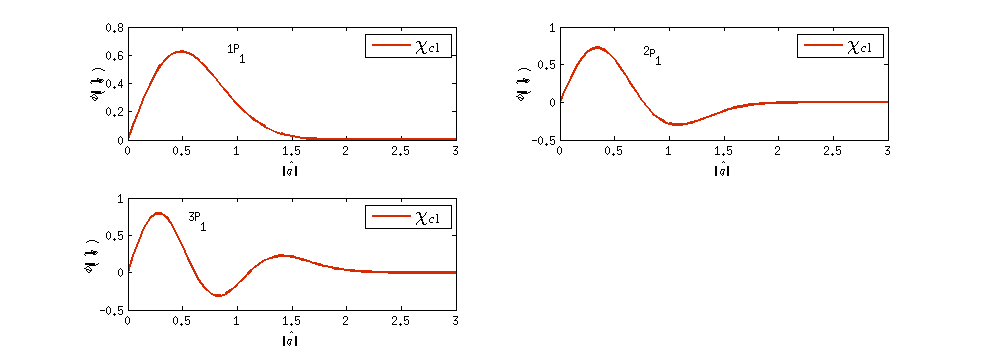}
\caption{Plots of wave functions for axial vector ($1^{++}$)
quarkonia $\chi_{c1}$ Vs $\widehat{q}$ (in Gev.) for the states
$1P, 2P$ and $3P$.}
\end{figure}

Now, the perturbative inclusion of the coulomb term will reduce
mass spectrum for axial vector meson in the same form as Eqs.(19)
-(21) for the scalar case, except that the inverse range parameter
$\beta_{s}$ has got to be replaced by $\beta_{A}$. This complete
mass spectral equation for ground and excited states of $1^{++}$
is:

\begin{equation}
E_{A}\phi_{A}(\hat{q})=[-\beta_{A}^{4}\overrightarrow{\nabla}^{2}_{\hat{q}}+\hat{q}^{2}+V_{coul}^{A}]\phi_{A}(\hat{q}),
\end{equation}

The solutions of the above spectral equation is:

\begin{equation}
\frac{1}{2\beta_{A}^2}\{\frac{M^{2}}{4}-m^{2}+\frac{\beta_{A}^{4}C_{0}}{\omega_{0}^{2}}\sqrt{1+2A_{0}(N+\frac{3}{2})}\}+
\gamma_{A} <V_{coul}^{A}>=(N+\frac{3}{2}); N=2n+l; n=0,1,2....,
\end{equation}

with $l=1$, where $<V_{coul}^{A}>$  is the expectation value of
$V_{coul}^{A}$ between the unperturbed states of given quantum
numbers $n$ (with $l=1$) for axial vector mesons, and has been
weighted by a factor of
$\gamma_{A}=\frac{C_{0}^2\beta_{A}^{4}}{(m_{1}+m_{2})^{2}}$ to have
the coulomb term dimensionally consistent with the harmonic term.
Its expectation values for the $1P, 2P$, and $3P$ states are:
\begin{eqnarray}
&&\nonumber
<1P|V_{coul}^{A}|1P>=-\frac{64\pi\alpha_{s}}{9\beta_{A}^{2}},\\&&
\nonumber
<2P|V_{coul}^{A}|2P>=-\frac{32\pi\alpha_{s}}{18\beta_{A}^{2}},\\&&
<3P|V_{coul}^{A}|3P>=-\frac{1856\pi\alpha_{s}}{213\beta_{A}^{2}}
\end{eqnarray}

From the mass spectral equation, one can see that, the mass
spectra again depends not only on the principal quantum number
$N$, but also the orbital quantum number $l$. We are now in a
position to calculate the numerical values for mass spectra of
heavy scalar quarkonia with the input parameters of our model as
mentioned above. The results of mass spectral predictions of heavy
equal mass scalar mesons for both ground and excited states with
the above set of parameters is given in table 2. The results on
masses of ground ($1P$) and excited ($2P$ and $3P$) states of
quarkonia, $\chi_{c1}$ are given in Table 5.

\begin{table}[hhhhh]
  \begin{center}
\begin{tabular}{|l|l|l|l|l|l|}
  \hline
   & BSE-CIA &Expt \cite{olive14} &Pot.Model\cite{godfrey85} &BSE\cite{wang10} &RQM\cite{ebert13}\\
   \hline
  $M_{\chi c1}(1p_1)$ & 3.5128& 3.510$\pm $0.0007 & 3.440 &  & 3.413 \\
   \hline
  $M_{\chi c1}(2p_1)$ & 3.9398 & 3.871.69$\pm$0.0017 &3.920  &3.928  &3.870  \\
  \hline
 $ M_{\chi c1}(3p_1)$ & 4.4616&  &  &4.228  & 4.301 \\
  \hline
  \end{tabular}
\caption{Mass spectrum of ground and excited states of $\chi_{c1}$
with quantum numbers $J^{PC}=1^{++}$ in GeV. units with the above
set of parameters.}
\end{center}
\end{table}

\section{Two-photon decays of scalar quarkonium}
We now study the two-photon decay width of a scalar quarkonium
($0^{++}$), which proceeds through the quark-triangle diagrams
shown in Fig.3.
\begin{figure}[h]
\centering
\includegraphics[width=12cm]{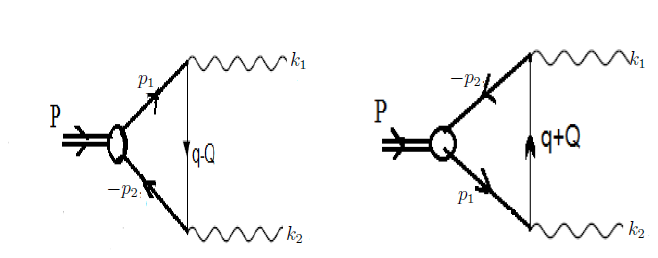}
\caption{Diagrams contributing to the two-photon decays of scalar
($0^{++}$) quarkonia}
\end{figure}

Let $P$ be the total momentum of the scalar quarkonia, and
$k_{1.2}$ be the momenta of the two emitted photons with
polarizations $\varepsilon_{1,2}$ respectively. Then we can write
$P=k_1+k_2$, and let $2Q=k_1-k_2$. The invariant amplitude for
this process can be written as:
\begin{equation}
M_{fi}(S\rightarrow\gamma\gamma)=\frac{i\sqrt{3}(ie_{Q})^{2}}{m^{2}+\frac{M^2}{4}}\int
\frac{d^{3}\widehat{q}}{(2\pi)^{3}}Tr[\Psi^{s}(\widehat{q})[\slashed{\epsilon_1}(m+i\slashed{Q})\slashed{\epsilon_2}+
\slashed{\epsilon_2}(m+i\slashed{Q})\slashed{\epsilon_1}]].
\end{equation}
where $e_Q=+\frac{2}{3}e$ for $c\overline{c}$, The 3D structure of
Dirac wave function, $\Psi^{s}(\widehat{q})$ is given in Eqs.(9),
and (17). The propagators for the third quark in the two diagrams
is expressed as, $S_F(q\mp Q)=\frac{-i(\slashed{q}\mp
\slashed{Q})+m}{(q\mp Q)^2+m^2}$. Now for heavy hadrons, where the
system can be regarded as non-relativistic, it is a good
approximation to take the internal momentum $q<<M$, and hence
$q^2<<Q^2$, where it can be seen that $Q^2=\frac{M^2}{4}$. Using
the propagator expressions given above, and evaluating trace over
the gamma matrices, we can write the invariant amplitude given
above as:
\begin{eqnarray}
&&\nonumber
M_{fi}(S\rightarrow\gamma\gamma)=(\epsilon_{1}.\epsilon_{2})F_{S},\\&&
F_{S}=(\frac{16\alpha_{em}}{3\sqrt{3}\pi^{2}})\frac{mM}{m^{2}+\frac{M^2}{4}}\int
d^3\widehat{q}\phi_s(\widehat{q})
\end{eqnarray}

where $F_{S}$ is the decay constant for scalar quarkonium,
$\chi_{c0}$. The decay width for the process can then be expressed
as,
\begin{equation}
\Gamma(S\rightarrow\gamma\gamma)=\frac{1}{32\pi M}|F_{S}|^{2}.
\end{equation}

\begin{table}[hhhhh]
  \begin{center}
\begin{tabular}{|l|l|l|l|l|l|}
  \hline
   & BSE-CIA &Expt \cite{olive14} &Pot.Model\cite{godfrey85} &BSE\cite{munz} &RQM\cite{ebert}\\
   \hline
  $\Gamma_{\chi_{c0}(1p_1)\rightarrow\gamma\gamma}$ & 2768.05& 2341.50 & 1290.00$\pm$3.45 &1390.00 & 2900.00 \\
   \hline
  $\Gamma_{\chi_{c0}(2p_1)\rightarrow\gamma\gamma}$ & 1548.20 &  &950.00$\pm$3.88  &1110.00$\pm$130.00  &1900.00  \\
  \hline
 \end{tabular}
\caption{Two-photon decay widths of ground and first excited
states of $\chi_{c0}(1P)$ and $\chi_{c0}(2P)$ in eV with the input
set of parameters given after Eq.(7)}
\end{center}
\end{table}

\section{Discussions} We have employed a 3D reduction of BSE (with
a $4\times 4$ representation for two-body ($q\overline{q}$) BS
amplitude) under Covariant Instantaneous Ansatz (CIA) for deriving
the algebraic forms of the mass spectral equations for scalar,
pseudoscalar, vector and axial vector quarkonia using the full BS
kernel comprising of the one-gluon-exchange and the confining part
in an approximate harmonic oscillator basis that led to analytic
solutions (both eigen functions and eigen values). We thus obtain
the mass spectra of $c\overline{c}$ quarkonia for ground and
excited states of $0^{++}$, $0^{-+}$ , $1^{--}$,and $1^{++}$
states. The mass spectral results for all these states which are
compared with the experimental data \cite{olive14}, and other
models for each state (where ever available), are given in Tables
2-5 respectively. The masses, and the algebraic forms of eigen
functions for each quarkonium state so obtained will be used for
calculating their various transitions.

Here we wish to mention that in principle, in $Q\overline{Q}$
quarkonia, the constituents are close enough to each other to
warrant a more accurate treatment of the coulomb term. Though for
$b\overline{b}$ systems, the  coulomb term will be extremely
dominant in comparison to confining term, however, it may not be
so unreasonable to treat coulomb term perturbatively for
$c\overline{c}$ systems. Here, we wish to point out that if we are
getting reasonable results for orbital $c\overline{c}$
excitations, it is mainly due to centrifugal effects\cite{mitra81}
which ensure that $c-\overline{c}$ separation is large enough to
feel the effect of confining term more strongly than the coulomb
term. Further, the present approach is on lines of some earlier
works \cite{mitra81,vinodkumar08,kalinovsky05} where the
OGE(coulomb) term is treated perturbatively for charmed mesons and
baryons. Similarly in a recent work \cite{vijayakumar13}, the 3D
harmonic oscillator wave functions were used as a trial wave
functions for obtaining $c\overline{c}$ spectrum and their decays,
where these trial wave functions did not take into account the
importance of coulomb potential for heavy quarkonium systems.
Similar treatment was earlier followed in \cite{godfrey90}.

Also, in our recent work \cite{hluf16} in which one of us (SB) was
involved, we did not take into account the coulomb interactions
between $c\overline{c}$, and $b\overline{b}$ states, and used only
the confining interaction to study their spectra and decays on
lines of other works \cite{koll,babutsidze,olsson}.

However, in the present approach, using confining interactions
alone, we first analytically derive the algebraic forms of wave
functions for various $c\overline{c}$ states, which are of
harmonic oscillator form. Though we then introduce coulomb term
perturbatively for $c\overline{c}$, (and not for $b\overline{b}$),
this present calculation is a substantial improvement over our
previous work in \cite{hluf16} (where we did not treat coulomb
interaction at all). However, for $b\overline{b}$ states, a more
exact non-perturbative treatment of coulomb interaction would be
needed.

Further, our results on $c\overline{c}$ mass spectra suggest that
perturbative incorporation of OGE with use of harmonic oscillator
wave functions derived analytically is closer to reality than some
of the previous approaches\cite{vijayakumar13,godfrey90} where
they used h.o. wave functions only as trial wave functions which
did not take into account the importance of coulomb potential for
heavy quarkonium systems.

All numerical calculations have been done using Mathematica. We
selected the best set of 6 input parameters (given after Eq.(11)),
that gave good matching with data for masses of ground and excited
states of $c\overline{c}$, and $b\overline{b}$ quarkonia for
pseudoscalar and vector states. The same set of parameters above
was also used to calculate the leptonic decay constants of
$\eta_{c}, \eta_{b}, J/\psi$, and $\Upsilon$, as well as the
two-photon and two-gluon decay widths of $\eta_{c}$, and
$\eta_{b}$ in \cite{hluf16}, which were in in reasonable agreement
with experiment and other models.

Using the same set of input parameters, we have predicted the full
mass spectrum of ground ($1P$), and excited ($2P$ and $3P$) states
of $\chi_{c0}$ and $\chi_{c1}$. Also we calculated the full
spectrum of $1S,...,4S$ states of $\eta_c$, as well as $1S,...,4D$
states of $J/\Psi$. The analytic forms of wave functions derived
for $0^{++}$ states of $c\overline{c}$ were employed to calculate
their two-photon decays .

We wish to also point out that the present calculation with
perturbative incorporation of the one-gluon-exchange interaction,
also lifts up the degeneracy between the $S$ and $D$ states of
vector quarkonia, and also giving a better agreement with the data
for these states. The results obtained for the ground ($1P$) state
of $\chi_{c0}$, as well as the ground and first excited states of
$\chi_{c1}$ are in reasonable agreement with data. However, there
are open questions about the quantum number assignments of the
state $X(3915)$, which is available in PDG tables. Some authors
\cite{guo,olsen} argue that it is difficult to assign $X(3915)$ to
$\chi_{c0}(2P)$, and that it could be $\chi_{c2}(2P)$. The
possible mass of $\chi_{c0}(2P)$ was recently predicted by
\cite{guo} as $3.837\pm 0.0115$ MeV. We then worked out the decay
widths of $\chi_{c0}$ for ground (1P), and excited (2P) states,
and compared our results with data and other models in Table 6. We
further calculated the two photon decay widths of $\chi_{c0}$ for
$1P$, and $2P$ states. Though our results are smaller than data
\cite{olive14}, but compares well with other models
\cite{godfrey85,munz}. Further a large variation in decay widths
can be seen in other models.

We further wish to point out that this analytic approach (giving
explicit dependence of spectra on principal quantum number $N$)
under heavy quark approximation gives a much deeper insight into
the spectral problem, than the purely numerical approaches
prevalent in the literature. The plots of the analytical forms of
wave functions for various $J^{PC}$ states, obtained as solutions
of their mass spectral equations are also are given in Figs.1- 2
for scalar and axial vector quarkonia. The correctness of our
approach can be gauged by the fact that these plots are very
similar to the corresponding plots of amplitudes for these
quarkonia in \cite{wang10} obtained by purely numerical methods.
The analytical forms of eigen functions for ground and excited
states so obtained can be used to evaluate the various other
processes involving scalar and axial vector quarkonia, which we
intend to do next.

\textbf{Acknowledgements}: This work  was carried out at Addis
Ababa University, Ethiopia, and at Chandigarh University, India.
The authors wish to thank both the institutions for the facilities
provided during the course of this work. One of us (LA) wishes to
thank Samara University, Ethiopia for support for his Doctoral
programme. He also wishes to thank Chandigarh University for the
hospitality during his visit during August-October 2017.

{99}
\end{document}